\documentclass[letterpaper]{article} 
\usepackage{aaai2026arxiv}  
\usepackage{times}  
\usepackage{helvet}  
\usepackage{courier}  
\usepackage[hyphens]{url}  
\usepackage{graphicx} 
\usepackage{natbib}  
\usepackage{caption} 
\usepackage{algorithm}

\usepackage[most]{tcolorbox}
\usepackage{tabularx}
\usepackage{booktabs}
\usepackage{multirow}
\usepackage{algpseudocode}

\usepackage{newfloat}
\usepackage{listings}
\DeclareCaptionStyle{ruled}{labelfont=normalfont,labelsep=colon,strut=off} 
\floatstyle{ruled}
\newfloat{listing}{tb}{lst}{}
\floatname{listing}{Listing}
\title{Judging by the Rules: Compliance-Aligned Framework for \\ Modern Slavery Statement Monitoring}
\author {
    Wenhao Xu\equalcontrib\textsuperscript{\rm 1,\rm 2},
    Akshatha Arodi\equalcontrib\textsuperscript{\rm 1},
    Jian-Yun Nie\textsuperscript{\rm 2},
    Arsène Fansi Tchango\textsuperscript{\rm 1}
}
\affiliations {
    \textsuperscript{\rm 1}Mila - Quebec AI Institute \quad
    \textsuperscript{\rm 2}Université de Montréal \\
    \{akshatha.arodi-nagaraja, arsene.fansi.tchango\}@mila.quebec, \{wenhao.xu, jian-yun.nie\}@umontreal.ca
}

\begin{document}

\maketitle

\begin{abstract}
Modern slavery affects millions of people worldwide, and regulatory frameworks such as Modern Slavery Acts now require companies to publish detailed disclosures. However, these statements are often vague and inconsistent, making manual review time-consuming and difficult to scale. While NLP offers a promising path forward, high-stakes compliance tasks require more than accurate classification: they demand transparent, rule-aligned outputs that legal experts can verify. Existing applications of large language models (LLMs) often reduce complex regulatory assessments to binary decisions, lacking the necessary structure for robust legal scrutiny. We argue that compliance verification is fundamentally a rule-matching problem: it requires evaluating whether textual statements adhere to well-defined regulatory rules. To this end, we propose a novel framework that harnesses AI for rule-level compliance verification while preserving expert oversight. At its core is the Compliance Alignment Judge (CA-Judge), which evaluates model-generated justifications based on their fidelity to statutory requirements. Using this feedback, we train the Compliance Alignment LLM (CALLM), a model that produces rule-consistent, human-verifiable outputs. CALLM improves predictive performance and generates outputs that are both transparent and legally grounded, offering a more verifiable and actionable solution for real-world compliance analysis.
\end{abstract}

\section{Introduction}

Modern slavery continues to affect over 50 million people worldwide~\cite{free2022global}. To address this urgent issue, several countries have enacted Modern Slavery Acts (MSAs), requiring companies to disclose how they assess and address slavery risks within their supply chains. While these laws have led to the publication of thousands of corporate statements each year~\cite{ukgovmodernslaveryregistry, amsa2018_register, canadianmodernslaveryact}, the disclosures vary widely in clarity, structure, and substance. The current reliance on fully manual review hinders large-scale enforcement, leaving substantial gaps that allow harmful practices to persist~\cite{ChambersVastardis2020}. With over 80,000 modern slavery statements published globally and limited capacity for expert review, enforcement remains a major challenge~\cite{bora2025aimscheck}. This gap presents a unique opportunity for AI-driven solutions that are engineered for demonstrable social benefit when guided by strong domain expertise and rigorous traceability. 

Natural Language Processing (NLP) offers a promising path toward scalable compliance monitoring. However, the interdisciplinary nature of this task, spanning legal interpretation, policy enforcement, and societal accountability demands capabilities beyond those of standard classification models~\cite{Chen2025BeyondYO, pistilli2023stronger, rinderle2023predictive}. Modern slavery statements often contain vague, unstructured and promotional language that blends compliance content with corporate messaging, complicating the extraction and verification of criteria~\cite{bora2025}, where conventional NLP approaches frequently fall short~\cite{Ariai2024NaturalLP}.

\begin{figure*}[t]
  \centering
  \includegraphics[width=\textwidth]{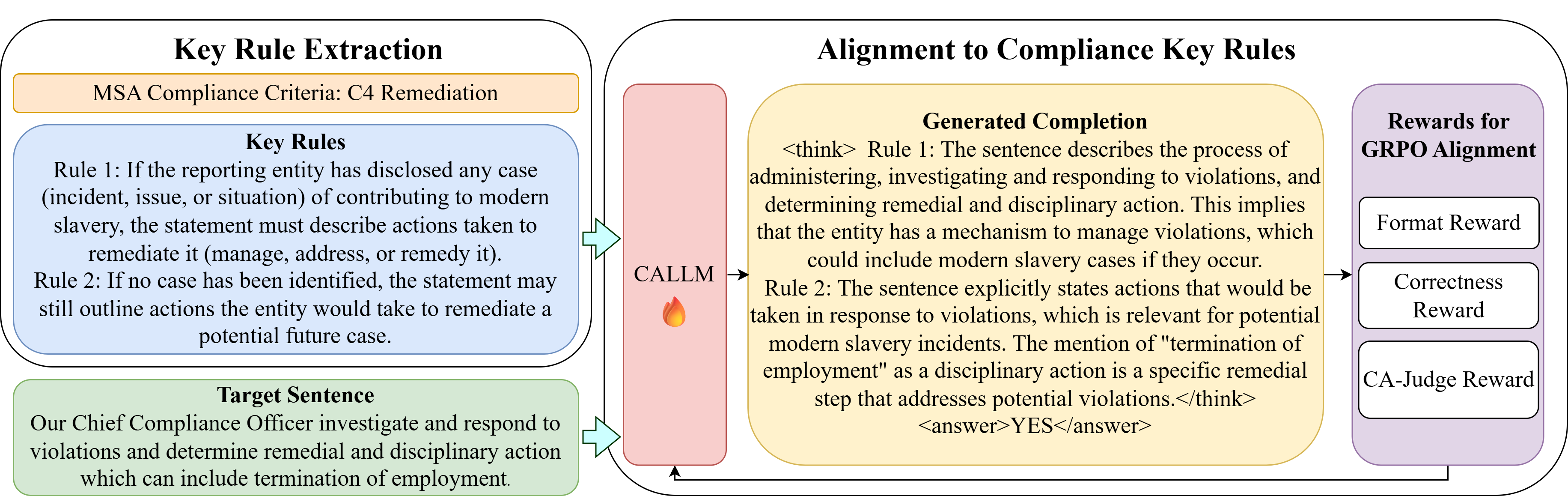}
  \caption{Overview of our framework, which consists of two steps: (1) \textit{Key Rule Extraction} derives natural language rubrics for a compliance criterion. (2) \textit{Alignment} trains CALLM to generate rule-aligned outputs using feedback from CA-Judge. The figure shows an example for the \textit{C4 Remediation} criterion under the Australian Modern Slavery Act, including the corresponding key rules, a target sentence to be classified as compliant (YES) or non-compliant (NO), and CALLM’s rule-aligned generation.}
  \label{fig:wide}
\end{figure*}

Effective compliance verification must prioritize not only accuracy but also traceability. For example, under the Australian Modern Slavery Act~\cite{australianmodernslaveryact}, the \textit{Approval} criterion demands explicit attribution to the principal governing body or CEO. Generic or vague language fails to meet this standard. Systems that rely on superficial cues risk costly errors: false negatives may unfairly penalize compliant firms, while false positives may allow violations to go undetected. In such high-stakes domains, outputs must provide transparent, rule-aligned justifications that enable human auditors to trace model outputs back to recognizable legal standards. Additionally, standard evaluation metrics such as precision and recall offer only aggregate insights and fail to capture whether individual regulatory criteria have been correctly justified~\cite{lipton2018mythos, doshi2017towards}. This limitation is especially critical in domains like compliance verification, where decisions often hinge on nuanced, criterion-specific judgments. As a result, in current practice, human reviewers remain essential for verifying whether disclosures meet statutory obligations~\cite{chaleshtori2024evaluating, zwickel2023compliance}. Consequently, prior work has recommended human-in-the-loop systems~\cite{bora2025, bora2025aimscheck}.

In this work, we propose a compliance-aligned framework and train the Compliance Alignment LLM (CALLM) for AI-assisted verification of modern slavery disclosures. Our approach is designed to produce rule-consistent outputs for human audit. At the center of our framework is a Compliance Alignment Judge (CA-Judge), a rule-aware evaluator that assesses whether the output of a model satisfies the statutory requirements of a given compliance criterion. Unlike general-purpose LLM-as-judge models~\cite{zheng2023judging, openai2023gpt4, korbak2023model}, CA-Judge is grounded in domain-specific regulatory logic. Its structured evaluations are intended to mirror how humans assess compliance: through attention to rule coverage, specificity, and clarity. CALLM uses the feedback of CA-Judge as a reward signal during training,  encouraging it to generate outputs that are explicitly aligned with relevant criteria rules (Figure~\ref{fig:wide}). Our goal is to verify whether this improves both task performance and the auditability of generated outputs. By generating rule-aligned assessments, CALLM aims to support faster, more transparent compliance checks.

Recent work has shown that chain-of-thought (CoT) reasoning does not reliably reflect model decision-making and should not be treated as an interpretability method~\cite{barez2025chain}. Instead of using model outputs to explain final decisions, we generate rule-aligned outputs intended for human verification and not as explanations to be trusted on their own. In our framework, the compliance verifier must review each rule-level rationale and make the final determination. Rather than self-justification, our goal is to generate rule-grounded outputs for expert audit at the criterion level. This human-in-the-loop, rule-aligned design supports oversight, addresses enforcement bottlenecks, and improves real-world deployability. Moreover, CALLM uses relatively small models (3B), promoting adoption and reproducibility.

We evaluate CALLM using both quantitative and qualitative metrics. It outperforms the baselines in both compliance classification and the rule-adherence of generated rationales. A human preference study confirms that outputs are better aligned with statutory rules, making them easier to verify. These results support our hypothesis: aligning model outputs with domain-specific rubrics improves both performance and usability. Further cross-jurisdictional analysis demonstrates that the framework generalizes effectively across jurisdictions.

Our core contribution lies in the training framework that leverages a compliance-specific CA-Judge to produce rule-grounded outputs. The novelty of our approach is in integrating regulatory rubrics (key rules derived from compliance statutes) directly into the training process, enabling more structured generation. While we focus on modern slavery disclosures, the proposed methodology is applicable to other regulatory domains where decisions depend on rule adherence. We release code and implementation guidelines to facilitate adoption and extension across domains~\footnote{\url{  https://github.com/mila-ai4h/aims-reasoning-alignment}}. We hope this work contributes to broader efforts in using AI for social good and inspires the community to engage more deeply with underexplored modern slavery compliance challenges.

\begin{figure*}[!ht]
\begin{tcolorbox}[
    colback=gray!10!white,
    colframe=gray!80!black,
    boxrule=0.3pt,
    left=2pt,
    right=2pt,
    top=1pt,
    bottom=1pt,
    enhanced jigsaw,
    sharp corners,
    fontupper=\fontsize{8pt}{10pt}\selectfont,
    boxsep=2pt
]

\textbf{KEY RULES for C4 Remediation:}

\textbf{Rule 1:} If one or more cases have been declared by the reporting entity where it caused or contributed to modern slavery, the statement should describe the actions taken to remediate these cases. \\
\textbf{Rule 2:} If no modern slavery case has been identified by the reporting entity, it may still describe actions used to remediate hypothetical cases if one should occur in the future.

\textit{Relevant examples:} Actions proposed by the reporting entity to remediate modern slavery cases. Corrective actions and sanctions to remediate modern slavery cases include, for example: conducting inquiries and investigations involving stakeholders. \\
\textit{Irrelevant examples:} Actions proposed to mitigate the risks of modern slavery instead of remediating existing cases. E.g., “We understand the importance of workers knowing their rights and addressing violations when necessary.”

\textbf{EVALUATION DIMENSIONS:}

1. \textit{Accuracy}: Identifies all relevant compliance gaps based on the key rules. Applies legal concepts exactly as defined—no misinterpretation or omission. No partial credit—any incorrect rule application makes the reasoning inaccurate. \\
2. \textit{Clarity}: Reasoning is logically structured with clear, step-by-step justification. Avoids vague terms, ambiguity, or unsupported claims. Final answer must clearly follow from the reasoning. \\
3. \textit{Fidelity to Key Rules}: All relevant key rules must be explicitly mentioned and addressed. Paraphrasing is allowed only if legal meaning is preserved. Irrelevant or external standards are penalized. \\
4. \textit{Consistency}: No internal contradictions; reasoning and conclusion must align. \\
5. \textit{Evidence Use}: Cites or paraphrases relevant rule clauses accurately. No new rules shall be introduced. \\
6. \textit{Cognitive Behaviors (Verification \& Reflection)}: Demonstrates explicit self-checking, cross-referencing, or reflection.

\end{tcolorbox}
\caption{Evaluation dimensions used by the CA-Judge. The key rules for the \textit{C4 Remediation} criterion under the Australian Modern Slavery Act are also shown, along with examples of relevant and irrelevant sentence types.}
\label{fig:evaluation_dimension}
\end{figure*}
\section{Related work}

\paragraph{Compliance Verification} Recent work has explored the use of NLP models for compliance classification, particularly in the regulatory domain~\cite{sun2025compliance,lore2023ai}.~\cite{bora2025} introduced a large-scale corpus of modern slavery statements from Australia and small evaluation sets for UK and Canada~\cite{bora2025aimscheck}. However, the models in these works typically focus on surface-level features and lack explicit reasoning in line with legal rules, limiting their utility in high-stakes applications. Similarly, approaches developed for the COLIEE competition~\cite{rabelo2022overview} are not suitable, as they emphasize cross-document retrieval or entailment. In contrast, compliance setting is normative: it requires verification of statutory criteria with traceable, rule-grounded justifications, redefining the decision target and evaluation goal.

\paragraph{Reasoning Models} Recent advances in reasoning, such as Chain-of-Thought prompting~\cite{wei2022chain}, self-consistency decoding~\cite{wang2022self}, and instruction-tuned LLMs like GPT-4~\cite{openai2023gpt4}, improve accuracy and explainability by encouraging intermediate steps. However, they often produce generic outputs and lack alignment with domain-specific rules, which is crucial in legal and compliance contexts. Recently, it is shown that these explanations frequently fail to reflect the model's actual internal reasoning process, creating illusions of interpretability, problematic in high-stakes applications~\cite{barez2025chain}.

\paragraph{Reinforcement Learning for Alignment}
Reinforcement Learning with Human Feedback (RLHF) is a widely adopted strategy for aligning LLMs with human preferences~\cite{ouyang2022training}. However, reward signals based on final-answer correctness are often too sparse, limiting training efficiency~\cite{Liu2024ImprovingMR}. Group Relative Policy Optimization (GRPO)~\cite{Shao2024DeepSeekMathPT} addresses this by comparing candidate outputs within groups, removing the need for an explicit reward model.

\paragraph{LLM-as-a-Judge}
The LLM-as-a-Judge framework leverages language models to evaluate outputs via preference-aligned feedback rather than rigid metrics~\cite{Zheng2023JudgingLW, Gu2024ASO}. While effective in open-ended reasoning tasks~\cite{Saha2025LearningTP}, these approaches typically operate in general domains without structured rules.

\paragraph{Our Contribution}
We unify reasoning-based generation, alignment, and judge feedback in a compliance setting, where explicit rules guide both training and evaluation, and the CA-Judge provides fine-grained supervision to ensure outputs are coherent and aligned with regulatory criteria.

\section{Dataset and Task}

We use the AIMS.au dataset~\cite{bora2025}, which contains annotated sentences from 5,731 modern slavery statements submitted by Australian companies in relation to the Australian Modern Slavery Act. 

The task is framed as sentence-level binary classification across multiple reporting criteria. These criteria differ in complexity. For example, \textit{Signature} is a simple criterion that checks if the document is signed, while \textit{C4 Mitigation} requires assessing whether a company has described concrete steps to mitigate modern slavery risks. We focus on 7 complex and 2 simple criteria in our experiments, selected based on the availability of expert-defined key rules for each. Each criterion is governed by a distinct set of key rules, defined by domain experts in~\cite{bora2025} based on the Australian Modern Slavery Act. Examples are shown in Figure~\ref{fig:evaluation_dimension}. Models are trained on the dataset’s training split and evaluated on the test split. To address class imbalance, we apply random downsampling to the training set.
(see Appendix~\ref{ap:dataset_details}).

Each training instance includes a target sentence, its surrounding context, and the key rules for a specific criterion. These are formatted into a structured prompt. The model is trained to generate a predicted label (\texttt{Yes}/\texttt{No}) along with a justification grounded in the provided rules, encouraging rule-aligned outputs. More details on the dataset and key rules and a full example prompt are shown in the Appendix Figure~\ref{fig:appendix:prompt_template_with_context}.

\section{Compliance Alignment LLM}
\label{sec:cv_model}
To align model reasoning with domain-specific rule requirements, we propose a rule-aligned training framework, shown in Figure~\ref{fig:wide}. This framework consists of two main stages: key rule extraction and alignment to these rules.

\subsection{Key Rule Extraction}

We translate compliance criteria into structured, rule-based rubrics that we define as key-rules. The key rules can be written by experts or extracted using LLMs and then reviewed. These rules distill regulatory knowledge into deterministic rubrics framed in natural language. Each rule clearly specifies what constitutes a valid versus invalid response. These serve as both training targets and evaluation anchors. Key rules of \textit{C4 Remediation} criterion under the Australian MSA is provided in Figure~\ref{fig:evaluation_dimension}.

\subsection{Alignment to Compliance Key Rules}

We train a model to align its outputs to the key-rules using feedback from the CA-Judge. To achieve this, we apply \textit{Group Relative Policy Optimization} (GRPO)~\cite{Shao2024DeepSeekMathPT}, a reward-based fine-tuning method that uses scalar scores, selected for its adaptability and efficiency in policy optimization, particularly in low-resource settings.

\subsection{Compliance Alignment Judge (CA-Judge)}
\label{sec:cv_judge}
\begin{figure}[t]
     \centering
     \includegraphics[width=\linewidth]{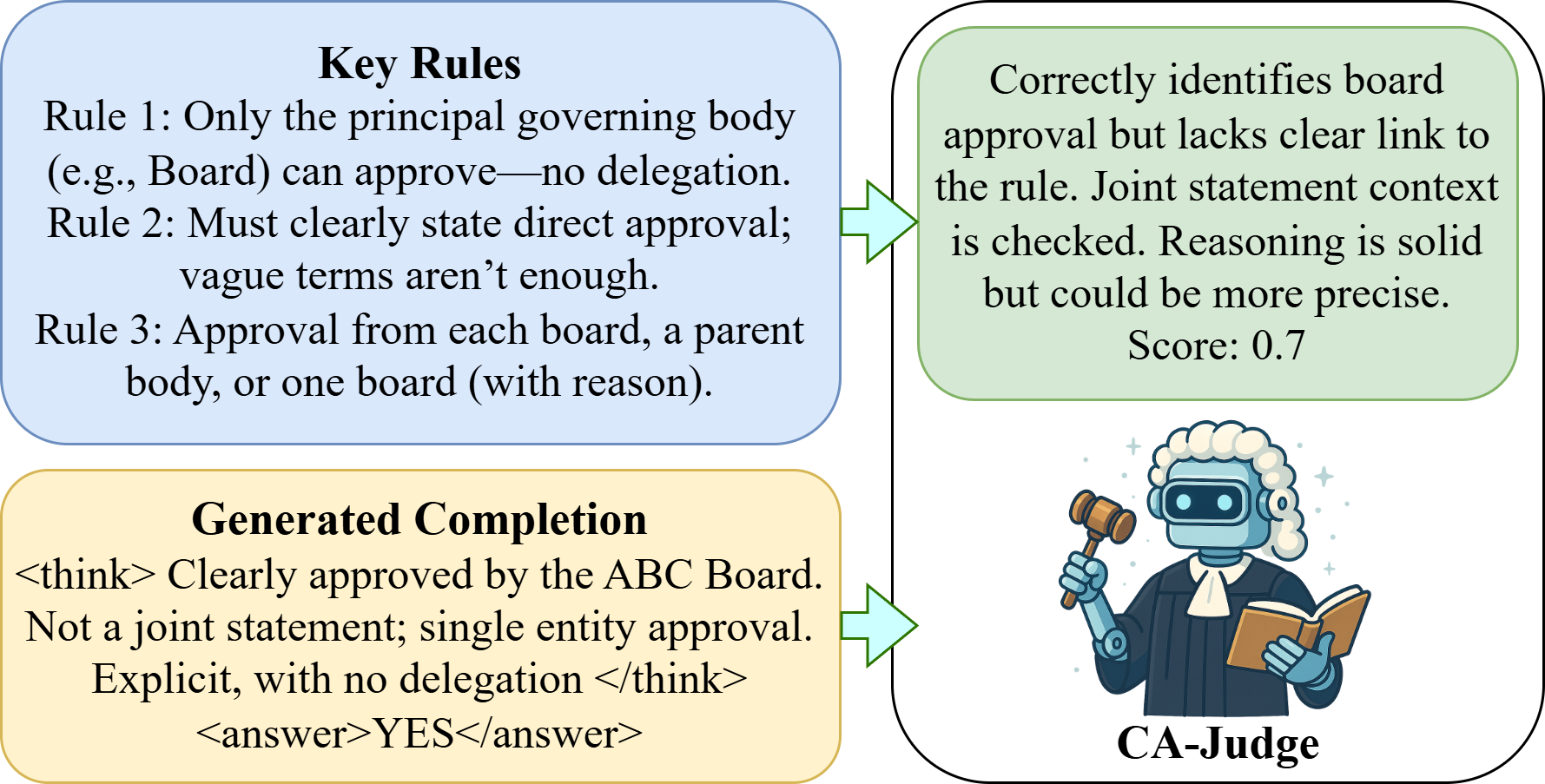}
     \caption{The Compliance Alignment Judge evaluates generated completion from a model against predefined key rules for compliance and generates a decision score with justification. The score reflects the degree of rule compliance and quality, enabling fine-grained, rule-aligned evaluation.}
     \label{fig:intro-figure}
\end{figure}

We introduce the CA-Judge as an evaluation method for compliance verification. CA-Judge employs a large language model to assess the alignment between a model’s outputs and a set of predefined rules. Given the model’s rationale, predicted label, and the associated key rules for a specific criterion, CA-Judge returns a scalar score that captures overall alignment across six evaluation dimensions (see Figure~\ref{fig:evaluation_dimension} for their definitions). The six dimensions target auditability (\textit{accuracy}, \textit{fidelity}), explainability (\textit{clarity}, \textit{evidence use}), and reliability checks (\textit{consistency}, \textit{verification}). It also produces a justification for the score. This approach provides rule-grounded feedback and closely mirrors how humans assess compliance, making it well suited for regulatory domains (see Figure \ref{fig:intro-figure}).

\subsection{Reward Design and Training Pipeline}  
We design a composite reward function to guide the fine-tuning of our CALLM model (base model and training details are in Section \ref{sec:experimental_setup}). The reward integrates surface-level formatting checks, prediction correctness, and alignment with key rules. During training, each instance includes input text, key compliance rules, and a gold label. The model generates multiple completions, each containing structured reasoning and a final prediction. These outputs are scored along three dimensions: surface-level fidelity, correctness, and rule-alignment. A total reward is computed as a weighted sum of these components, and GRPO is used to rank completions and update model parameters. The training regime is shown in Algorithm~\ref{alg:grpo_training}.

\subsubsection{Surface-Level Fidelity}  
To encourage structurally correct outputs, we apply two rewards~\cite{Shao2024DeepSeekMathPT}:

\paragraph{1. Format Match.}  
A binary reward indicating whether the output matches the expected format:
\begin{equation}
R_{\text{format}}(i) = \mathbf{1}\left\{\, \text{match}(c_i) \,\right\}
\end{equation}
where $c_i$ is the model’s completion.

\paragraph{2. XML Tag Count.}  
Rewards the presence of required tags while penalizing excess length:
\begin{align}
R_{\text{xml}}(i) &= \min\left[\,1,\; \max\left[\,0,\; \right. \right. \notag \\
& \xi_r \sum_{t \in \mathcal{T}} \mathbf{1}\{ t \in c_i \} - \xi_p\, \Delta_i\, ]\, ]
\end{align}
where \( \mathcal{T} \) is the set of required tags, \( \Delta_i \) is excess character count. $\xi_r = \frac{1}{\lvert \#\mathcal{T} \rvert}$ is the reward for required tags. $\xi_p$ is the penalty for excess text.

\subsubsection{Correctness.}  We assign a binary reward based on whether the model's predicted label matches the gold label:
\begin{equation}
\label{r_corr}
    R_{\text{corr}}(i) = \mathbf{1}\left\{\, \hat{y}_i = y_i \,\right\}
\end{equation}

\subsubsection{Rule-alignment.}  
The Compliance Alignment Judge provides a scalar score that evaluates the alignment of the model's reasoning and final prediction with the key-rules:
\begin{equation}
\label{r_judge}
    R_{\text{judge}}(i) = \text{CAJudge}(\text{rules}, \text{reasoning}, \hat{y}) \in [0, 1]
\end{equation}

\subsubsection{Total Reward.}  
We define the total reward as a weighted sum of the above components:
\begin{align}
R_{\text{total}}(i) =\; & \lambda_1 \cdot R_{\text{format}}(i) + \lambda_2 \cdot R_{\text{xml}}(i) \notag\\
& + \lambda_3 \cdot R_{\text{corr}}(i) + \lambda_4 \cdot R_{\text{judge}}(i)
\end{align}

\begin{algorithm}[ht]
\caption{Rule-Aligned Training}
\label{alg:grpo_training}
\begin{algorithmic}[1] 
\State \textbf{Input:} Dataset $\mathcal{D}$ with (context, key rules, label), LLM $f_\theta$, CA-Judge, weights $\lambda_i$
\State policy model $\pi_\theta\!\leftarrow\!\pi_{\text{init}}$
\While{not converged}
  \State Sample batch $\{(x_i, r_i, y_i)\}_{i=1}^B \sim \mathcal{D}$
  \ForAll{$x_i$ in batch}
    \State Generate completions $\{c_i^{(1)}, \ldots, c_i^{(K)}\}$ using $f_\theta$
    \ForAll{$c_i^{(k)}$}
      \State Get: $r_{\text{format}}$, $r_{\text{xml}}$, $r_{\text{correct}}$, $r_{\text{judge}}$
      \State $r_{\text{total}} = \sum_{j=1}^{4} \lambda_j \cdot r_j$
    \EndFor
    \State Rank completions by $r_{\text{total}}$ and compute GRPO loss
  \EndFor
  \State Update model parameters $\theta$; $\pi_{\text{old}}\!\leftarrow\!\pi_\theta$
\EndWhile
\State \textbf{Return:} Fine-tuned model $f_\theta$
\end{algorithmic}
\end{algorithm}

\section{Base Models \& Experimental Setup}
 \label{sec:experimental_setup}
We evaluate the proposed framework against a range of models as baselines, spanning Zero-Shot, Few-Shot and Fine-tuned settings. We include GPT-4o in zero/few-shot configurations using Chain-of-Thought prompts. In few-shot settings, we use 3 examples that were randomly selected from a diverse list of real-world cases to maximize coverage across typical compliance scenarios. Additionally, we evaluate DeepSeek-R1 and its distilled variant, DeepSeek-R1-Distill-Qwen-7B~\cite{deepseek_r1_distill_qwen_7b}, in zero-shot settings. We also include Pre-Trained and Fine-tuned variants of the CALLM base model (ablations). We do a full evaluation one criterion at a time for all 9 criteria.

\paragraph{CA-Judge:} We use JudgeLRM-7B~\cite{chen2025judgelrmlargereasoningmodels}, a judgment-oriented LLM trained to be good at judging tasks, as our Compliance Alignment Judge, responsible for scoring the alignment between model-generated justifications and the compliance key rules (see Section~\ref{sec:cv_judge}). We selected JudgeLRM-7B, as it has fast inference speed and is trained in judgment tasks and has strong performance in judging-based evaluation benchmarks, consistently outperforming general-purpose models on such tasks. An example of the scoring rubric and full prompts are 
in the Appendix.

\paragraph{CALLM:} We use Qwen2.5-3B-Instruct~\cite{yang2025qwen} as the base model, chosen for its strong performance in instruction tuning and generation efficiency on compliance-style prompts. The model is fine-tuned using our framework as described in Section \ref{sec:cv_model}. In our experiments, we use equal weights for all rewards: \( \lambda_1 = \lambda_2 = \lambda_3 = \lambda_4 = 1 \). We set all $\lambda$ values to 1 as fixed training hyperparameters, following the prior work~\cite{shao2024deepseekmathpushinglimitsmathematical, deepseekai2025deepseekr1incentivizingreasoningcapability, dao2025alphamazeenhancinglargelanguage}. We set  $\xi_r$ to 0.25 as there are 4 required tags, to make the rewards fall in [0,1] and $\xi_p$ to 0.001.

\paragraph{Evaluation Metrics:} We report F1 score as the primary metric for classification performance. To assess the rule adherence of model justifications, we additionally assess the model outputs using CA-Judge score, which reflects the degree to which model outputs align with compliance key rules. We opted against reference-based metrics as they correlate poorly with human judgments, and require high-quality gold reasoning references, which are difficult to obtain in compliance settings without substantial expert annotation. We leverage CA-Judge that evaluates across multiple axes such as accuracy, clarity, and correctness, which is seen in recent works~\cite{liu2023gevalnlgevaluationusing, he2024socrevallargelanguagemodels, li2024llmsasjudgescomprehensivesurveyllmbased}. We also conduct human evaluation.
\section{Results}

\begin{table*}[ht]
\centering
\resizebox{\textwidth}{!}{%
\small
\begin{tabular}{llccccccc}
\toprule
Group & Criterion & \multicolumn{2}{c}{GPT-4o} & \multicolumn{2}{c}{DeepSeek} & \multicolumn{2}{c}{Base model} & CALLM \\
&   & ZS & FS & R1 & Distill & PT & FT & (Ours) \\
\midrule
\# Params & & 1800B & 1800B & 671B & 7B & 3B  & 3B & 3B \\
\midrule
\multirow{2}{*}{General}
    & Approval      & \textbf{0.855} & 0.843 & 0.837 & 0.464 & 0.349 & 0.755 & 0.786 \\
    & Signature     & 0.409 & 0.636 & 0.250 & 0.154 & 0.422 & 0.524  & \textbf{0.692} \\
\midrule
\multirow{3}{*}{C2}
    & Structure     & 0.619 & 0.658 & \textbf{0.678} & 0.350 &  0.310 & 0.535  & 0.572 \\
    & Operations    & 0.529 & \textbf{0.651} & 0.537 & 0.290 &  0.181 & 0.596  & 0.632 \\
    & Supply Chains & 0.420 & 0.556 & 0.399 & 0.260 &  0.192 & 0.550  & \textbf{0.601} \\
\midrule
\multirow{1}{*}{C3}
    & Risk Description & 0.422 & 0.450 & 0.495 & 0.260 &  0.237 & 0.564  & \textbf{0.712} \\
\midrule
\multirow{2}{*}{C4}
    & Mitigation    & 0.709 & 0.664 & 0.700 & 0.243 &  0.451 & 0.714  & \textbf{0.749} \\
    & Remediation   & 0.552 & \textbf{0.601} & 0.529 & 0.159 &  0.225 & 0.397  & 0.570 \\
\midrule
\multirow{1}{*}{C5}
    & Effectiveness & \textbf{0.518} & 0.492 & 0.504 & 0.342 &  0.267 & 0.394  & 0.439 \\
\midrule
\multirow{1}{*}{Overall}
    & (macro) & 0.559 & 0.617 & 0.548 & 0.280 &  0.293 & 0.559  & \textbf{0.639} \\
\bottomrule
\end{tabular}
}
\caption{F1 Scores Across Compliance Criteria for Baseline Models and our CALLM model. Best score in each row is highlighted. ZS = Zero-shot, FS = Few-shot, PT = Pre-trained, FT = Fine-tuned. Base model here indicates the base model of CALLM without CA-Judge supervision. Overall,  CALLM outperforms the baselines.}
\label{tab:baseline-comparison}
\end{table*}

\paragraph{Quantitative Analysis}Table~\ref{tab:baseline-comparison} presents F1 scores across the nine compliance criteria for a range of baseline models and our proposed model, CALLM. Despite having only 3B parameters, CALLM achieves the highest overall macro-F1 score (0.639), outperforming much larger models such as GPT-4o \footnote{Estimated at 1800B parameters} and DeepSeekR1 (671B). To ensure that performance gains are not solely due to fine-tuning, we include a fair baseline, ``base model FT'', identical in architecture and training data to CALLM but lacking our alignment feedback with CA-Judge. CALLM consistently outperforms this baseline, highlighting the value of our compliance-aligned optimization. CALLM shows particularly strong performance on challenging criteria such as\textit{ C2 Supply Chains}, \textit{C3 Risk Description}, and \textit{C4 Mitigation}, where it requires interpreting complex and sometimes subjective compliance rules. These results demonstrate the benefits of our alignment strategy, especially in tasks that demand structured rule-grounded reasoning. Interestingly, while CALLM excels in complex criteria, its performance on simpler ones like \textit{Approval} is slightly lower than GPT-4o. Our error analysis reveals that CALLM is stricter in rule adherence, and is sensitive to data issues—such as broken target sentences that omit key terms like ``approved by'', which are required by the rule definitions (see the Appendix for examples Figure \ref{fig:model_output_failure_1}).
Overall, CALLM strikes a favorable balance between model size and performance, surpassing all baselines under comparable or even more favorable resource settings.

We evaluate model outputs using CA-Judge to assess whether training with compliance-aligned feedback leads to outputs that better reflect the intended regulatory rules. While final qualitative validation is done with human evaluation, this automated assessment serves as a key intermediate check to validate our training framework on a held-out test set. In our setting, as expected, training with CA-Judge feedback leads to reasoning that better aligns with the compliance rules, specific to the task of regulatory rule alignment. Specifically, we compare CALLM to the second-best model from Table~\ref{tab:baseline-comparison}, GPT-4o with Few-shot chain-of-thought prompting. Across all nine criteria, CALLM consistently outperforms GPT-4o, achieving a substantially higher overall average score (0.74 vs. 0.53).
These results suggest that CALLM generates justifications that are more grounded in the key compliance rules, while GPT-4o often produces generic reasoning with limited alignment to compliance rubrics. CALLM shows particularly strong gains on challenging criteria such as C2 Structure, C3 Risk Description, and C4 Remediation, where precise rule application is both complex and critical. These results highlight the value of our framework guided by compliance-specific evaluations. Detailed results are in the Appendix.

\begin{figure*}[!ht]
\begin{tcolorbox}[
    colback=gray!10!white,
    colframe=gray!80!black,
    boxrule=0.3pt,
    left=2pt,
    right=2pt,
    top=1pt,
    bottom=1pt,
    enhanced jigsaw,
    sharp corners,
    fontupper=\fontsize{8pt}{10pt}\selectfont,
    boxsep=2pt
]

\textbf{CALLM Success Case for the C4 Remediation criterion} \\
\textbf{Sentence:}
...contributing to society in both developed markets as well as the developing regions of the world where we operate this is enshrined in our Code. \textbf{\textit{Our Chief Compliance Officer and our Regional Ethics Counsels administer our Code, investigate and respond to violations and determine remedial and disciplinary action which can include termination of employment.}} We have committed to conducting annual training on the Code for our employees...

\textbf{CALLM Reasoning:}
1. The target sentence clearly states that the Chief Compliance Officer (CCO) and Regional Ethics Counsels (RECs) administrate a Code, investigate and respond to violations, and determine remedial and disciplinary actions, which can include termination of employment.  
2. This describes a remediation action for modern slavery cases by specifying what actions can be taken (termination of employment).  
3. The sentence describes a process, which includes administering a Code, investigating and responding to violations, and determining remediation and disciplinary measures, aligning with Rule 1 since it refers to existing or potential cases.  
4. Even though it mentions the possibility of future cases, the actions outlined are clearly remediation and handling of existing cases (Rule 1 requirements specifically addressing potential future cases as well as existing ones). These actions are directly specified for Rule 2.  
5. The actions described are specific and concrete, providing clarity on what remediation measures the entity will take.

\textbf{CALLM Final Answer:} \textit{yes} (Ground truth: yes)

\textbf{CA-Judge response:} The model’s reasoning aligns well with the key rules. It correctly identifies that the sentence describes concrete remediation actions, such as termination, satisfying Rule 1. It also notes these actions could apply to future cases, covering Rule 2. The explanation is clear and well-structured but could be improved by more explicitly linking the actions to specific aspects of each rule. The model's final answer of "YES" is correct because the reasoning clearly supports that the described actions meet the criteria for remediation actions as per the key rules. \textit{Score: [[0.9]]}

\vspace{2pt}\hrule\vspace{4pt}

\textbf{CALLM Failure Case for the Approval criterion} \\
\textbf{Sentence:}
This Statement was approved by the boards of the three reporting entities covered by this Statement... Busways Gosford Pty Ltd. Busways North West Pty Ltd. \textbf{\textit{Busways Pacific Pty Ltd.}} Busways South Australia Pty Ltd...

\textbf{CALLM Reasoning:}
The key rule is whether the statement shows approval by the principal governing body. The sentence \textit{"This Statement was approved by the boards of each of the reporting entities covered by this statement"} appears to meet this.  
1. Approval Authority: Specifies approval by the board of each reporting entity, indicating direct involvement of governing bodies.  
2. Direct Approval: No indirect terms like “on behalf of” or “considered by.”  
3. Verification: “The Board of Busways Pty Ltd” confirms direct, explicit board-level approval.  
The sentence clearly states board-level approval for each entity, with no mention of delegation or committees.

\textbf{CALLM Final Answer:} \textit{yes} (Ground truth: no)

\textbf{CA-Judge response:} The reasoning identifies board approval but misses ambiguity in “boards” and fails to consider joint statement alternatives. It only partially aligns with the key rules, leading to an incorrect answer.  
\textit{Accuracy:} Notes board approval but overlooks that “boards” may imply a group rather than a singular principal body.  
\textit{Clarity:} Logically structured but misses ambiguity in language and does not address required alternatives.  
\textit{Fidelity to Key Rules:} Skips conditions for joint approval—e.g., higher body or justification for single board.  
\textit{Consistency:} Internally consistent but incomplete in rule coverage.  
\textit{Evidence Use:} Cites rules but is vague on “boards” vs. principal body.  
\textit{Cognitive Behaviors:} Some reflection shown, but alternative scenarios are not explored.  
\textit{Final Answer:} Incorrect due to incomplete handling of joint approval rules.  
\textit{Score: [[0.5]]}

\end{tcolorbox}
\caption{
Illustrative use cases of CALLM with CA-Judge scoring. \textit{Sentence} includes the target sentence (bold) and context. \textbf{Top:} CALLM correctly predicts compliance for C4 Remediation. CA-Judge assigns a high score, reflecting strong alignment with key rules (shown in Figure~\ref{fig:evaluation_dimension}) and well-structured reasoning. \textbf{Bottom:} CALLM incorrectly predicts compliance for Approval. CA-Judge detects flaws in the reasoning, correctly lowering the score in line with the rule violations (rules in Figure~\ref{fig:intro-figure}).
}
\label{fig:model_output_success_failure}
\end{figure*}

\paragraph{Qualitative Analysis}
\begin{figure}[t]
  \centering

  \includegraphics[width=0.8\linewidth]{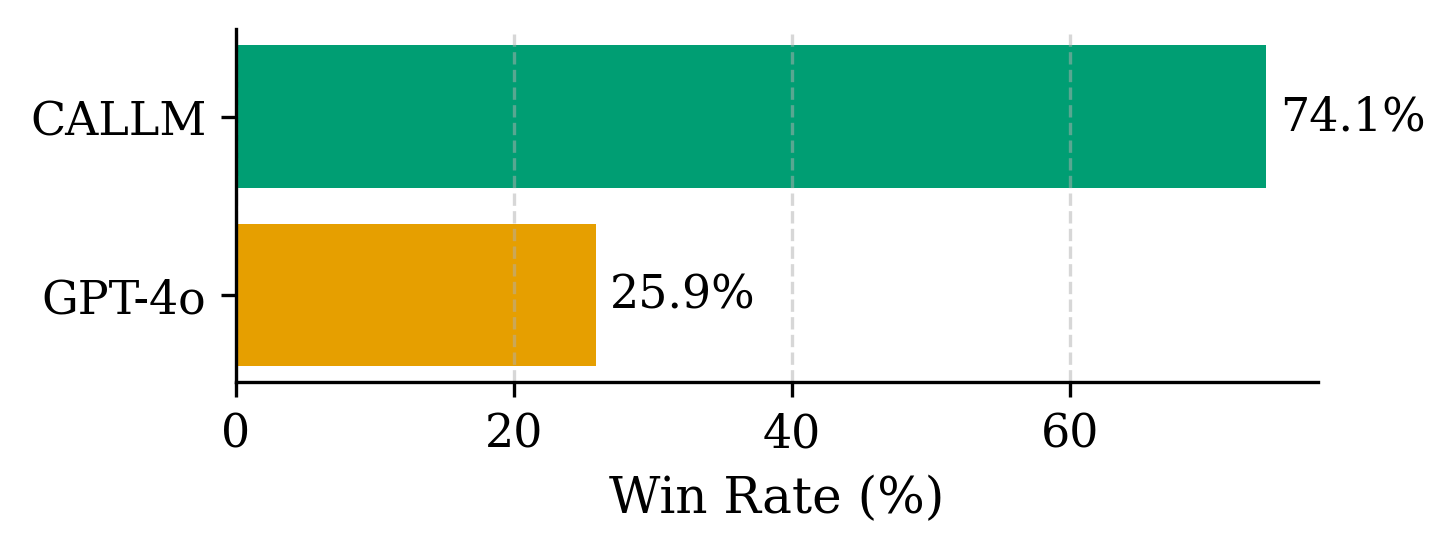}
  \caption{Human‐preference comparison between our model and the baseline.}
  \label{fig:human-pref-bar}
\end{figure}

To assess the quality of model-generated justifications, we conducted a human study comparing CALLM and GPT-4o. The study involved 270 total responses from five volunteer participants (two with prior knowledge of modern slavery compliance and three without). For each of the nine compliance criteria, we randomly sampled six representative examples (three compliant, three non-compliant), excluding low-quality or overly short sentences. Both models were prompted using the same setup to generate structured rule-aligned outputs and a final answer. Responses were uniformly formatted and randomly assigned as Option A or B to ensure blinding.

Participants were asked to choose which response better satisfied the compliance rules, using the same evaluation dimensions as the CA-Judge: accuracy, clarity, fidelity to key rules, and consistency. Across 54 paired comparisons, CALLM was preferred in 74.1\% of cases, receiving 200 preference votes compared to 70 for GPT-4o, as shown in Figure~\ref{fig:human-pref-bar}. CALLM was especially favored in criteria requiring nuanced compliance interpretation (e.g., distinguishing corporate structure from operational involvement). To further assess alignment between human and CA-Judge evaluations, we compared preference direction per criterion. In 7 out of 9 criteria (77.8\%), human judgments matched the CA-Judge’s scoring direction, demonstrating strong agreement between the two. More details are in the Appendix~\ref{ap:human_survey}.

\paragraph{Ablation Study} Table~\ref{tab:baseline-comparison} includes ablations of our CALLM model. We compare: (1) the pre-trained base model (\textbf{PT}), (2) a model fine-tuned using surface-level rewards for format and label correctness (\textbf{FT}), and (3) our full model, CALLM, trained with all the rewards, including rule-aligned feedback from the Compliance Alignment Judge (\textbf{CALLM}). CALLM achieves the highest overall performance, with improvements on complex criteria such as \textit{C3 Risk Description}, and \textit{C4 Mitigation}. These results demonstrate that fine-tuning based on format and correctness improves upon the base model, but the greatest gains are achieved by aligning reasoning with key rules from the CA-Judge.

\section{Discussion}

\paragraph{Case Studies}
We present qualitative case studies demonstrating that our training framework yields rule-aligned justifications. Figure~\ref{fig:model_output_success_failure} (top) shows an example. We observe that in several instances, baseline models predicted the correct label but either omitted justification or provided vague rationales unrelated to the key compliance rules.
In contrast, CALLM generates explanations that explicitly referenced relevant risk-related criteria (e.g., identifying when descriptions failed to specify industries, regions, or supply chain elements associated with modern slavery). 
Additional detailed cases are shown in the Appendix.

\paragraph{Error Analysis} 
We examine failure cases of CALLM. As shown in Figure~\ref{fig:model_output_success_failure} (bottom)  CALLM incorrectly predicts compliance. Note that the CA-Judge correctly identifies the rule violations and assigns a lower score.
In some cases, the model generates verbose or generic reasoning that nominally matches key-rules but lacks specificity. Ambiguous inputs such as implicit approvals or vague governance language remain challenging, suggesting the need for tighter rule definitions or hybrid supervision strategies. 

\begin{table}[th]
\centering
\small
\begin{tabular}{lcccc}
\toprule
\textbf{Model} & \textbf{Params (B)} & \textbf{AU} & \textbf{UK} & \textbf{CA} \\
\midrule
GPT-4o ZS CoT & 1800 & 0.559 & 0.500 & 0.560 \\
GPT-4o FS CoT  & 1800 & 0.617 & 0.573 & 0.614 \\
DeepSeek-R1          & 671  & 0.548 & 0.505 & 0.550 \\
Base Model FT        & 3    & 0.559 & 0.560 & 0.589 \\
\textbf{CALLM (ours)}& \textbf{3} & \textbf{0.639} & \textbf{0.620} & \textbf{0.617} \\
\bottomrule
\end{tabular}
\caption{Models are trained on AU and evaluated on AU, UK and CA held-out sets show generalization.}
\label{tab:xjur}
\end{table}

\paragraph{Cross-jurisdiction Generalization} 
To test generalization beyond Australia, we evaluated CALLM on modern slavery statements from the UK and Canada~\cite{bora2025aimscheck}. CALLM is trained on the AU train data and evaluated on the test sets of AU, UK and CA. We use the same setup as Table~\ref{tab:baseline-comparison}. CALLM consistently outperforms larger models and the fine-tuned baseline (macro-F1), yielding robust cross-jurisdictional performance (see Table~\ref{tab:xjur}).

\paragraph{Facilitation of follow-up work} We release code, prompt templates, and implementation guidelines to support real-world adoption and enable the community to build on our work whether by improving model performance, enhancing evaluation methods, or extending the framework to other compliance domains. Additional details, including the experimental setup, hyperparameters, limitations of our work, and ethical considerations, are provided in the Appendix.

\paragraph{Future Work} We aim to support multi-hop legal reasoning and finer-grained rule signals, further improving framework utility for complex policy and compliance challenges. 

\section{Conclusion}
We introduce a novel framework for aligning large language models with compliance requirements in high-stakes domains. Our approach centers on CA-Judge, which evaluates model outputs against key statutory criteria, and trains the Compliance Alignment LLM (CALLM) using a rule-aligned reward signal. We find that CALLM, by generating outputs that explicitly reference relevant rules, outperforms baseline models in both predictive accuracy and human preference. This is intended to enhance human-in-the-loop verification of modern slavery statements by providing rule-based justifications, to enable faster and more reliable review. By supporting rule-aligned reasoning at scale, we aim to reduce manual review burdens, increase accountability, and build trust in real-world deployments. We hope this work encourages broader integration of AI-assisted review in modern slavery compliance and encourages the AI community to join the global fight against modern slavery.

\section*{Acknowledgments}
We gratefully acknowledge Yuchen Hui for constructive and insightful discussions during the early stages of this project, and Ziyan Wang, Pedro Ferraz, Hager Radi, and Jeremy Pinto for their valuable suggestions throughout the experimental phase. This research was supported by compute resources provided by Mila (mila.quebec).

\appendix

\section{Limitations}

While our framework improves compliance alignment, several limitations remain. First, the effectiveness of our method depends on the quality and coverage of the rule rubric. Incompletely specified or ambiguous rules may lead to inconsistent reward signals from the CA-Judge. Second, our use of GRPO introduces sensitivity to batch composition: if all candidate outputs are weak, the model may reinforce suboptimal patterns. We adopted GRPO because its group‑relative advantage removes the need for a separate value network, making it more memory‑ and compute‑efficient in low‑resource settings. In contrast, Proximal Policy Optimization (PPO) \cite{schulman2017proximal} requires training an additional value model and extensive tuning, which exceeded our computational budget. Direct Preference Optimization (DPO) \cite{wu2024beta} was infeasible  due to lack of human preference data for the compliance domain. Although GRPO aligns well with our constraints, the relative performance of PPO and DPO in scenarios where ample human preference data and computational resources are available remains an open question that warrants systematic investigation. Third, although our CA-Judge provides dense supervision, it is still an approximation of expert judgment and may introduce bias or hallucination under adversarial prompting. Finally, we focus primarily on English-language compliance documents and modern slavery reporting; generalization to multilingual or other regulatory domains is an open direction for future work.

Due to constraints in computational resources, we were unable to train larger models (e.g., 7B or 14B parameters). We anticipate that integrating our framework with larger models could yield performance improvements, as scaling model size has been shown to enhance readability and informativeness in language tasks. Future work will focus on optimizing our framework to support larger models effectively.

Our preference survey for qualitative analysis was limited in scale and did not include domain experts, reflecting practical constraints in recruiting specialized annotators. While our findings suggest alignment with expert-like judgments, future work should involve a broader pool of reviewers, including compliance verification professionals.

\section{Ethical considerations}
This work focuses on improving the traceability and reliability of compliance classification models in high-stakes domains such as modern slavery reporting. While our methods are intended to support transparency and accountability, several ethical risks must be acknowledged. First, the CALLM and CA-Judge, like any LLM, may reflect biases from its pretraining data, potentially misjudging reasoning that deviates from dominant linguistic or institutional norms. Second, the automated assessment of legal compliance should not replace expert legal review, especially in ambiguous or contested cases. Our framework is designed to assist, not substitute, human oversight. Finally, any deployment of such models must be accompanied by safeguards to prevent misuse, such as automating regulatory approval or unfairly penalizing entities based on misclassified outputs. Future work should explore fairness auditing, stakeholder involvement, and domain-specific calibration to mitigate these concerns.

Our study includes a volunteer survey which involves only anonymous volunteer ratings of the model-generated outputs. No personal or sensitive information was collected during the process. Participation was entirely voluntary, with no compensation provided. Prior to conducting the survey, we thoroughly reviewed the content of the model outputs to ensure they did not contain any deceptive elements or sensitive topics. All materials were reviewed in advance to ensure minimal risk. As such, the study was conducted in line with standard ethical practices for low-risk survey.

\section{Experimental settings}
\label{ap:experiments_settings}
We release our code and experimental scripts\footnote{\url{https://github.com/mila-ai4h/aims-reasoning-alignment}} to facilitate reproducibility under the Apache License 2.0. We build on the implementation of the open-source notebook by \cite{schmid2024grpo} and make substantial modifications and enhancements to tailor it to our specific research objectives.

\paragraph{Settings \& Hyperparameters} Our experiments were conducted under the help of DeepSpeed \cite{10.1145/3394486.3406703}, Transformers \cite{Wolf2019HuggingFacesTS}, TRL \cite{vonwerra2022trl}, vLLM \cite{Kwon2023EfficientMM} libraries. 
The hyperparameter settings adopted during our experiments are detailed in Table \ref{tab:hyperparams}. All other parameters follow the default settings provided by the GRPOConfig\cite{vonwerra2022trl}.

\begin{table}[ht]
\centering
\begin{tabular}{lc}
\toprule
\textbf{Hyperparameters} & \textbf{Values} \\
\midrule
\texttt{learning\_rate} & $5e-6$ \\
\texttt{per\_device\_train\_batch\_size} & 8 \\
\texttt{gradient\_accumulation\_steps} & 1 \\
\texttt{lr\_scheduler\_type} & cosine \\
\texttt{warmup\_ratio} & 0.1 \\
\texttt{beta} & 0.04 \\
\texttt{max\_prompt\_length} & 4000 \\
\texttt{max\_completion\_length} & 1000 \\
\texttt{num\_generations} & 4 \\
\bottomrule
\end{tabular}
\caption{Hyperparameter settings used in our experiments.}
\label{tab:hyperparams}
\end{table}

Our training procedure consists of two phases:
\begin{itemize}
\item \textbf{Phase 1:} We applied supervised fine-tuning (SFT) with GRPO with a combination of reward functions focusing on format and correctness. This phase involved training for a maximum of 2000 steps to obtain the FT model.
\begin{align}
R_{\text{phase1}}(i) =\; & \lambda_1 \cdot R_{\text{format}}(i) + \lambda_2 \cdot R_{\text{xml}}(i) \notag\\
& + \lambda_3 \cdot R_{\text{corr}}(i)
\end{align}

\item \textbf{Phase 2:} Building upon the FT model, we resumed training by introducing the CA-Judge component, continuing for an additional 5000 steps to develop our final CALLM model.
\end{itemize}
\begin{align}
R_{\text{phase2}}(i) &=\;  \lambda_1 \cdot R_{\text{format}}(i) + \lambda_2 \cdot R_{\text{xml}}(i) \notag\\
& + \lambda_3 \cdot R_{\text{corr}}(i) + \lambda_4 \cdot R_{\text{judge}}(i)
\end{align}

For Phase 1, each criterion was trained using 2 NVIDIA A100 80GB GPUs, with the process taking approximately 3-4 hours. In Phase 2, training each criterion utilized 3 NVIDIA A100 80GB GPUs (1 addtional for the CA-Judge model), requiring about 40 hours to complete. All the 2 phases were conducted with BF16 and TF32 enabled.

To ensure consistency between training and evaluation, we employed the same temperature and other hyperparameters during evaluation as used in training (See Table \ref{tab:generation_hyperparams}). Addtionally, we set the random seed to 42 to maintain reproducibility.

\begin{table}[ht]
\centering
\begin{tabular}{lc}
\toprule
\textbf{Generation Hyperparameters} & \textbf{Values} \\
\midrule
\texttt{top\_k} & 50 \\
\texttt{top\_p} & 1.0 \\
\texttt{temperature} & 0.9 \\
\bottomrule
\end{tabular}
\caption{Generation hyperparameters aligned between training and evaluation phases.}
\label{tab:generation_hyperparams}
\end{table}

\begin{figure}[t]
    \centering
    \includegraphics[width=0.85\linewidth]{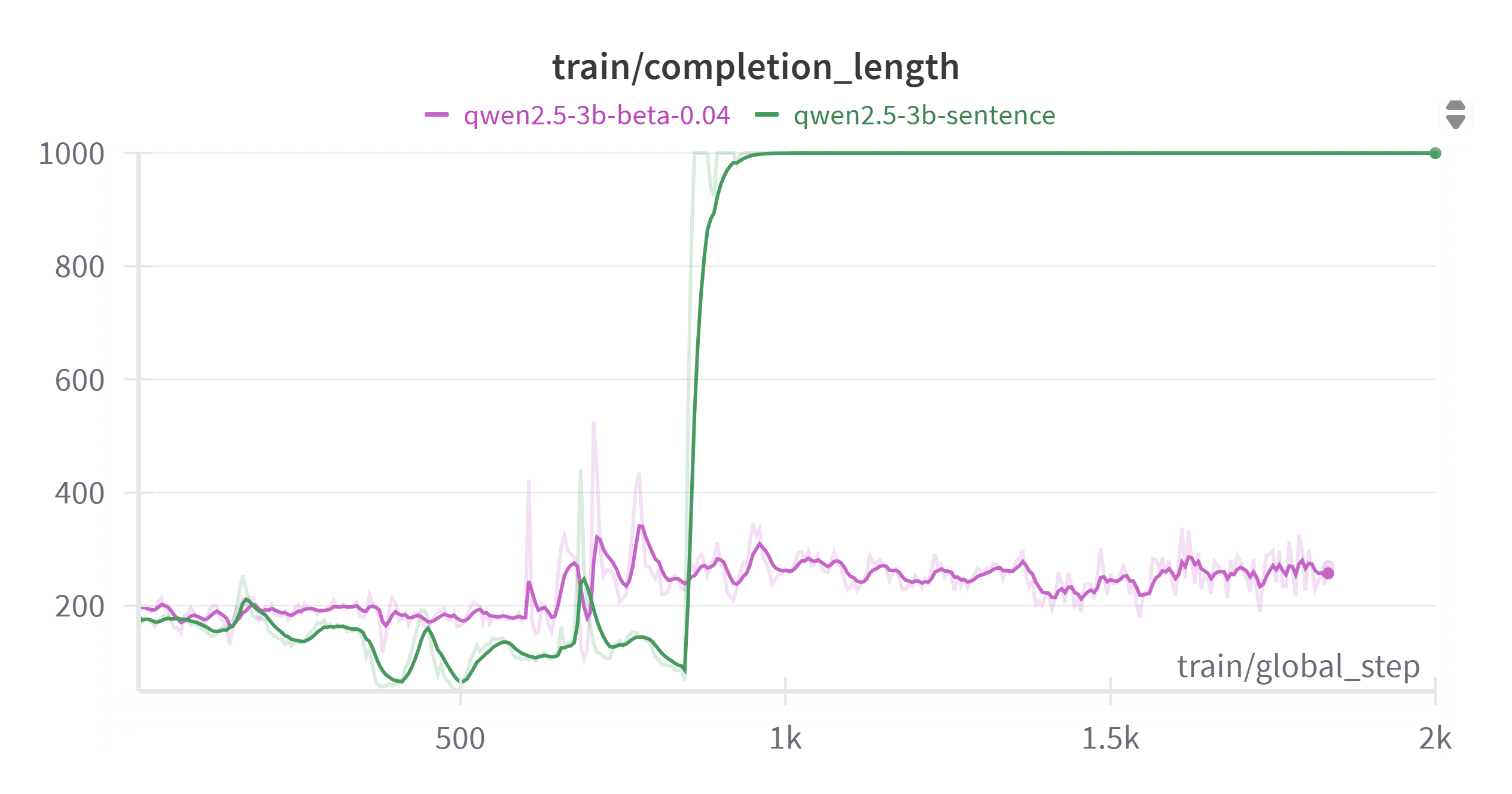}
    \caption{
        Completion lengths during GRPO training with different KL regularization strengths. 
        The green curve (\texttt{qwen2.5-3b-sentence}) corresponds to a low $\beta$ value ($\beta=0.001$), resulting in uncontrolled reward hacking and excessively long completions. 
        The purple curve (\texttt{qwen2.5-3b-beta-0.04}) uses a higher $\beta=0.04$, leading to stable completion lengths and effective constraint adherence.
    }
    \label{fig:beta_ablation}
\end{figure}

\paragraph{Reward hacking control} Although verifiable rewards are theoretically less susceptible to reward hacking than reward models learned from preference data, recent work shows that an inadequately regularised policy can still exploit loopholes in the constraint set \citep{Huang2024TheNI, Lambert2024TLU3P}. In our preliminary runs with GRPO we set the KL–divergence weight to a small value ($\beta=10^{-3}$) and repeatedly observed a failure mode in which the policy maximised the verifiable reward by generating extremely long, low–quality completions—classic reward-hacking behaviour. Raising the regularisation strength to $\beta=4\times10^{-2}$ (value from \cite{Shao2024DeepSeekMathPT}) aligned the policy much more closely with the reference model and virtually eliminated these pathological trajectories (Figure \ref{fig:beta_ablation}). This practical finding is consistent with the theoretical analysis of \cite{Mroueh2025ReinforcementLW}, who argue that sufficiently large KL penalties shrink the effective optimisation landscape and curb uncontrolled exploration. Our results therefore highlight the KL coefficient as a \emph{primary} hyper-parameter for mitigating reward hacking in GRPO; we recommend reporting its value and performing a principled sensitivity study whenever verifiable-reward RL is used.

\paragraph{GRPO limitations} We also find that GRPO's reliance on relative ranking makes it sensitive to group composition—if all completions in a batch are weak, the model may reinforce suboptimal patterns.

\section{Dataset Details}
\label{ap:dataset_details}
The dataset is highly imbalanced (Table \ref{tab:imbalance}), reflecting real-world reporting patterns, and includes both crowd-annotated and expert-annotated subsets to support reliable model evaluation.
\begin{table}[t]
\centering
\small
\setlength{\tabcolsep}{4.5pt}
\begin{tabular}{lccr}
\toprule
\textbf{Tasks} & \textbf{P(0)} & \textbf{P(1)} & \textbf{0:1 Ratio} \\
\midrule
Approval               & 99.11\% & 0.89\%  & 112:1 \\
Signature              & 99.14\% & 0.86\%  & 115:1 \\
C2 (structure)         & 94.04\% & 5.96\%  & 16:1  \\
C2 (operations)        & 90.93\% & 9.07\%  & 10:1  \\
C2 (supply chains)     & 93.07\% & 6.93\%  & 13:1  \\
C3 (risk description)  & 93.04\% & 6.96\%  & 13:1  \\
C4 (risk mitigation)   & 79.96\% & 20.04\% & 4:1   \\
C4 (remediation)       & 99.27\% & 0.73\%  & 135:1 \\
C5 (effectiveness)     & 94.50\% & 5.50\%  & 17:1  \\
\bottomrule
\end{tabular}
\caption{Label distribution and imbalance ratio (0:1) for each classification task.}
\label{tab:imbalance}
\vspace{1mm}
\raggedright
\footnotesize\textit{Each task contains on average $\sim$454,775 examples.}
\end{table}

\paragraph{Pre-processing}
\label{sec:data-prep}

We first discard empty or malformed sentences, yielding a clean corpus for the AIMS.au splits.  
To curb reward hacking caused by extreme class imbalance, we apply \emph{random downsampling} only to the training set, down-scaling the majority class until the ratio of positive to negative examples is $1{:}1$.  
The validation and test sets retain the original skew to ensure a realistic evaluation.

Downsampling is a lightweight alternative to more involved balancing schemes. While the RL community typically tackles imbalance with reward re-weighting \cite{Lin2019DeepRL}, balanced sampling alone has been shown to improve minority-class coverage and reduce policy bias in sequential decision problems \cite{Jiang2023OfflineRL,Liu2024PriorfreeBR}. Conversely, purely cost-weighted rewards become brittle under ratios $\!>\!$50:1; previous works \cite{Buda2017ASS,Shi2023HowRH} also report that weighting without resampling yields negligible or unstable gains, echoing classic findings in imbalanced-learning surveys \cite{krawczyk2016learning}.  

\subsection{GPT-4o and DeepSeek-R1}
We experiment with GPT-4o and DeepSeek-R1. We build upon the experiments conducted by \cite{bora2025}, leveraging the best-performing configurations identified in their study, GPT-4o with context of 100 words. We employ the same prompt structure as in their experiments. To further refine our approach, we conduct additional experiments by incorporating Chain of Thought (CoT) reasoning while maintaining the same context-based setup. These experiments were performed under both Zero-shot learning (no examples) and Few-shot learning (with a three to four examples per criteria). Finally, we replicate the same experiments using DeepSeek-R1, maintaining identical prompt structures to ensure comparability between models. GPT-4o model experiments were conducted using the API from \cite{openai2023gpt4}. Each criterion took between 8h to 15h to run, depending on the prompt, with a total of about \$1000 spent for the experiments. We setup 2.51bit quantized version of Deepseek-R1  640B using the implementation from \cite{unslothdeepseek}. The model was deployed using 4 H100 with 80 GB of memory using PyTorch and llama.cpp server. Each criteria took about 36 hours. 

\section{CA-Judge evaluation}
\begin{table}[!ht]
\centering
\label{tab:qwen-ablation}
\begin{tabular}{lcc}
\toprule
Criterion & GPT4o & CALLM \\
\midrule
Approval     & 0.39 & \textbf{0.69} \\
Signature    & 0.68 & \textbf{0.83} \\
C2 Structure  & 0.45 & \textbf{0.83}  \\
C2 Operations & 0.55 & \textbf{0.63}  \\
C2 Supply Chains  & 0.59 & \textbf{0.78}  \\
C3 Risk Description & 0.61 & \textbf{0.85} \\
C4 Mitigation   & 0.57 &  \textbf{0.69} \\
C4 Remediation  & 0.28 & \textbf{0.68}  \\
C5 Effectiveness    & 0.63 & \textbf{0.72} \\
\midrule
Overall  & 0.53 & \textbf{0.74} \\
\bottomrule
\end{tabular}
\caption{As expected, training with CA-Judge feedback results in rule-aligned reasoning in our experiments, as demonstrated on a held-out test set.}
\label{tab:cv_judge_results}
\end{table}

We evaluate model outputs with CA-Judge to automatically verify whether compliance-aligned training improves rule adherence. On a held-out test set, CALLM achieves higher overall average score (0.74 vs. 0.53) compared to GPT-4o (the second-best model from Table~\ref{tab:baseline-comparison}) across all nine criteria as shown in Table \ref{tab:cv_judge_results}. CALLM shows particularly strong gains on challenging criteria such as \textit{C2 Structure}, \textit{C3 Risk Description}, and \textit{C4 Remediation}, where precise rule application is both complex and critical.

\section{Survey to study human preference}
\label{ap:human_survey}
\begin{figure}[t]
    \centering
    \includegraphics[width=\linewidth]{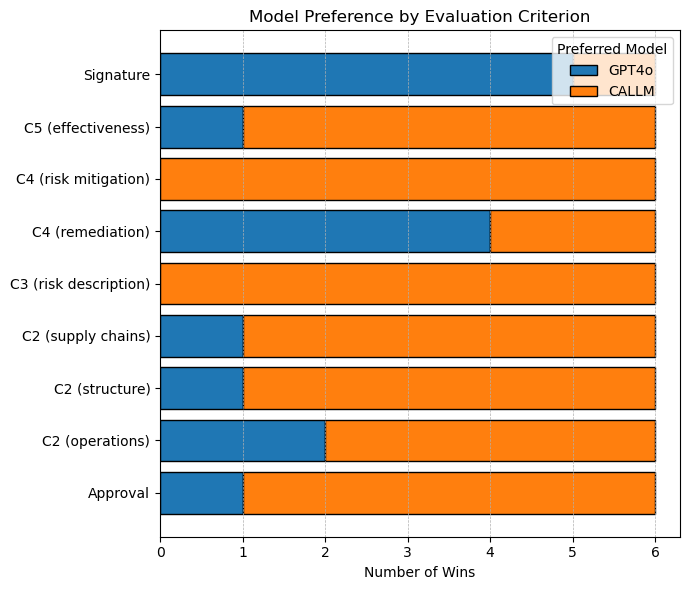}
    \caption{Model preference per criterion. Participants preferred CALLM in 7 out of 9 criteria.  
    }
    \label{fig:model_preference_criterion}
\end{figure}

\begin{figure*}[t]
    \centering
    \includegraphics[width=\linewidth]{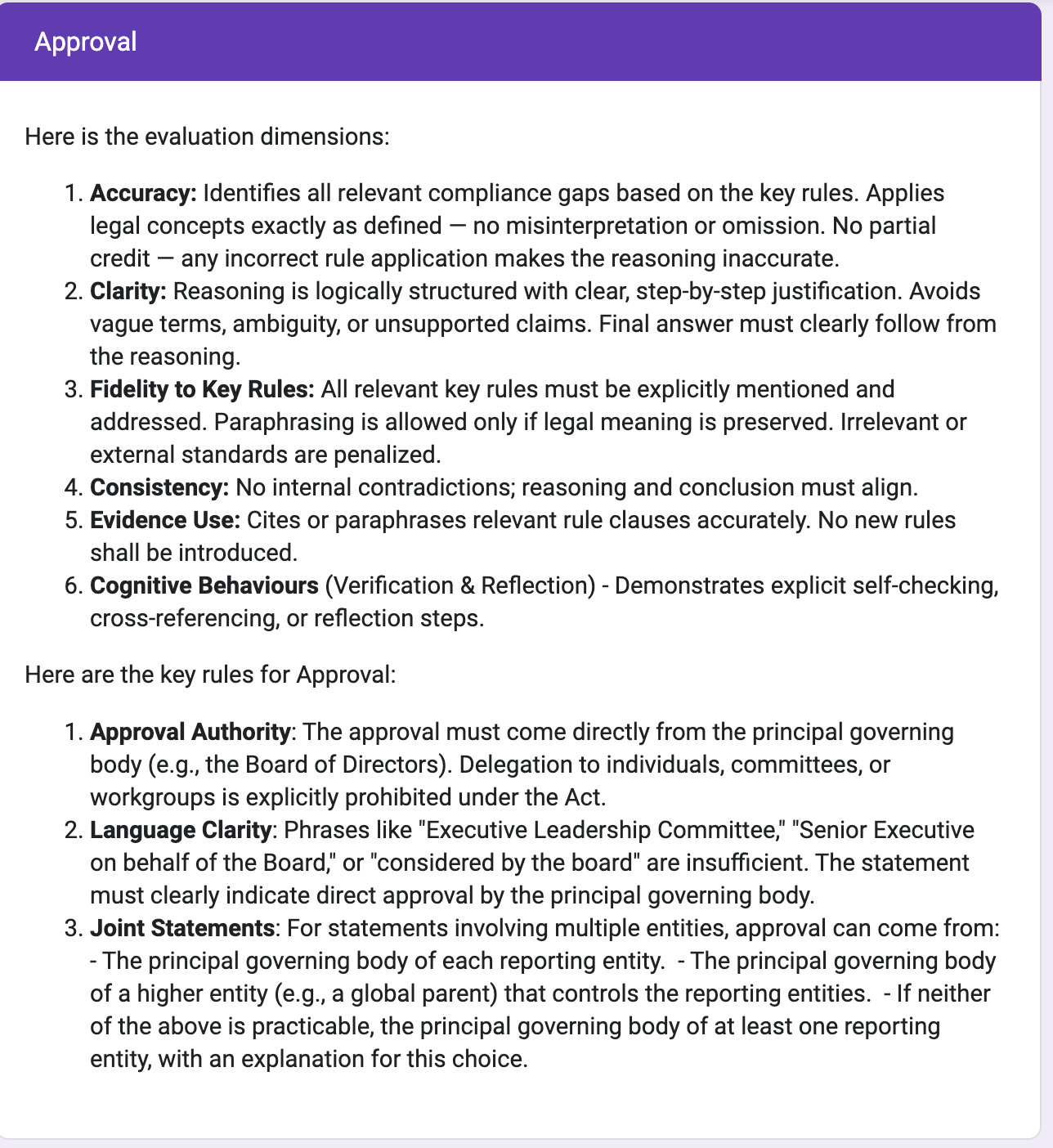}
    \caption{Survey interface explaining the task.  
    }
    \label{fig:human_survey_1}
\end{figure*}

\begin{figure*}[t]
    \centering
    \includegraphics[width=0.8\textwidth]{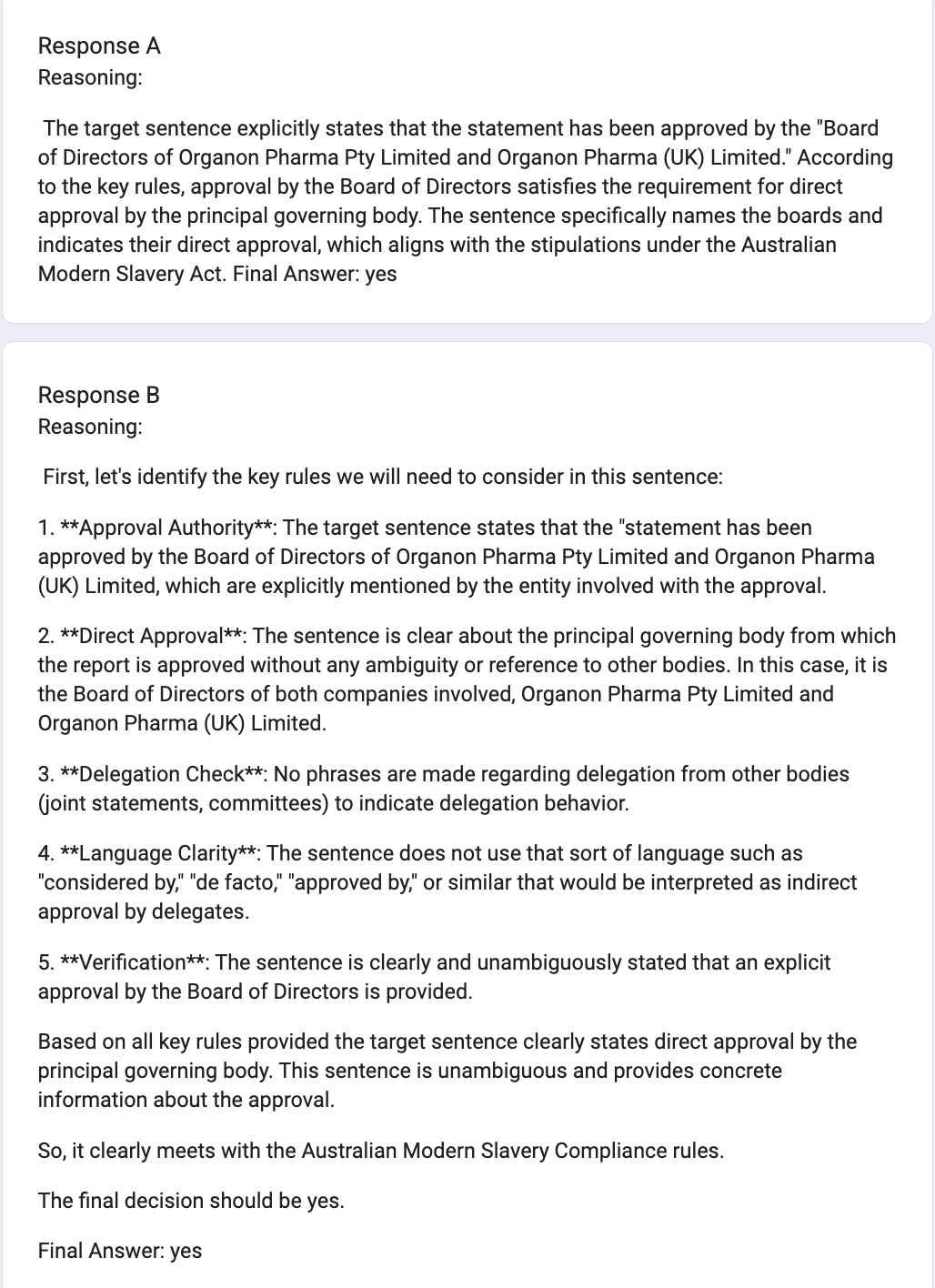}
    \caption{Survey with preference options.  
    }
    \label{fig:human_survey_2}
\end{figure*}

\paragraph{Protocol}
To assess how well model-generated outputs align with human preferences, we conducted a small-scale evaluation using a blind comparison setup. We recruited five volunteer participants, all based in North America, with graduate-level education and strong English proficiency. Two participants had prior familiarity with modern slavery compliance, while the other three did not. The study was conducted via an anonymous survey that presented participants with 54 examples, each showing a target sentence and two model-generated responses (CALLM vs. GPT-4o), randomized in order. For each example, participants were asked to choose which response better satisfied the compliance criterion. Figure \ref{fig:model_preference_criterion} shows model preference per criterion. No personal information was collected, and participants were not compensated. The full task took less than one hour to complete and involved binary preference judgments over 270 total responses.Given the minimal risk, anonymity, and non-sensitive nature of the task—along with the fact that no identifying or behavioral data were collected. The study followed common practice for low-risk, volunteer-based human evaluations in NLP. The survey interface is in Figures \ref{fig:human_survey_1} and \ref{fig:human_survey_2}.

\paragraph{Judge-Human Agreement.} To assess alignment between automated and human evaluations, we compute directional agreement across criteria based on whether both CA-Judge and human preferences favor the same model (CALLM or GPT-4o). Table~\ref{tab:judge-human-direction} shows per-criterion agreement: in 7 out of 9 cases (77.8\%), both judge and human evaluators align. Notably, disagreements occur in subjective criteria such as \textit{C4 Remediation} and \textit{Signature}, where humans sometimes favor surface-level fluency over rule fidelity

\begin{table}[t]
\centering
\small
\begin{tabular}{lcc}
\toprule
\textbf{Criteria} & \textbf{Judge} & \textbf{Human} \\
\midrule
Approval                    & 1 & 1 \\
C2 (operations)             & 1 & 1 \\
C2 (structure)              & 1 & 1 \\
C2 (supply chains)          & 1 & 1 \\
C3 (risk description)       & 1 & 1 \\
C4 (risk mitigation)        & 1 & 1 \\
C4 (remediation)            & 1 & -1 \\
C5 (effectiveness)          & 1 & 1 \\
Signature                   & 1 & -1 \\
\bottomrule
\end{tabular}
\caption{Comparison of model preference direction between CA-Judge and human evaluators across compliance criteria. A value of 1 indicates preference for CALLM, and -1 indicates preference for GPT-4o.}
\label{tab:judge-human-direction}
\end{table}

\section{Prompts}
\label{ap:prompts}
We provide the prompt templates used in our experiments for reference. Figure \ref{fig:appendix:prompt_template_with_context} illustrates the general prompt template applied during both training and evaluation across all models. Here we present the template for \textit{C3: Risk Description}, while we have 9 templates with differing key rules and examples for each criterion.

Figure \ref{fig:appendix:judge_prompt} displays the universal compliance alignment prompt used for CA-Judge. This template is employed both in training and during the reasoning quality evaluation (Table 2 in paper). 

\begin{figure*}
\begin{tcolorbox}[colback=gray!10!white, fontupper=\small, colframe=gray!80!black, title={Few-Shot Prompt Template (C3, Risk Description)}]
{\fontsize{8pt}{7pt}\selectfont
You are an analyst specialising in the review of modern slavery declarations made by Australian reporting entities under the Australian Modern Slavery Act (the Act).\newline\newline
Task:\newline
You will be given a target sentence and its surrounding context. Your goal is to determine whether the target sentence is compliant(relevent) based on the key rules. Compliance is determined solely by the content of the target sentence, not its context. Your answer must be YES or NO.
If a sentence is fragmented (e.g., missing words) but forms a complete compliant action when combined with the immediately preceding or following sentence, treat it as compliant.
Show your reasoning in \textless think\textgreater  \textless /think\textgreater  tags. And return the final answer in \textless answer\textgreater  \textless /answer\textgreater  tags. Before finalizing your answer, critically reflect: Have you rigorously cross-checked the statement(target sentence) against the key rules? Verify that no implicit assumptions or ambiguous phrasing were overlooked in your analysis.\newline\newline

Your task is to identify sentences in these declarations that describe the modern slavery risks identified by reporting entities.\newline

\#\#\# Key Rules:\newline
**Relevant sentences** Relevant sentences may include descriptions such as:\newline
Describing specific modern slavery risk areas such as geographic regions (e.g., Indonesia), industries (e.g., electronics), commodities (e.g., palm oil).\newline
Sentences describing any past or current instances of modern slavery within the entity's operations or supply chains.\newline
Offering an explanation for why they believe that their risks of modern slavery are low.\newline
**Irrelevant sentences** may include:\newline
Hypothetical scenarios or vague claims like “Modern slavery might exist within the technology sector,” as well as broad definitions of risk that do not specifically relate to the organisation’s operations or supply chains or its own and controlled entities.\newline
Descriptions of other business risks (e.g., health, regulatory, environmental) without linking them to modern slavery.\newline
Merely stating that the company has zero or low risks is too vague; they must clarify why and on what analysis this is based.
\newline\newline

\#\#\# Examples with reasoning:\newline\newline
Example 1:\newline
Target Sentence: "Whilst the majority of our products are supplied from Japan, we recognise that certain sectors and industries that we supply from, such as agriculture, are globally recognised as high risk industries."\newline\newline
\#\#\#\# Question: Is the target sentence relevant? (YES/NO)\newline\newline
\textless think\textgreater The sentence identifies specific industries (agriculture) that are high risk for modern slavery, linking it to the entity’s supply chain operations. This meets the key rules.\textless /think\textgreater \newline
\textless answer\textgreater YES\textless /answer\textgreater \newline

Example 2:\newline\newline
Target Sentence: "We procure goods and services from trusted suppliers across the world."\newline\newline
\#\#\#\# Question: Is the target sentence relevant? (YES/NO)\newline\newline
Reasoning: While this sentence mentions procuring goods and services from suppliers globally, it is too vague and lacks specific details about the suppliers, their products or services, locations, or other attributes that describe the supply chain.\newline\newline
Final Answer: NO\newline\newline

Example 3:\newline\newline
Target Sentence: "We continue due diligence across all our direct and indirect suppliers.”\newline\newline
\#\#\#\# Question: Is the target sentence relevant? (YES/NO)\newline\newline
Reasoning: This sentence mentions due diligence efforts but does not provide specific descriptions of suppliers, their products or services, locations, or other attributes. It lacks the detailed information required to describe the supply chain.\newline\newline
Final Answer: NO\newline\newline

The target sentence to classify is the following:
\newline
------------
\newline
\texttt{TARGET\_SENTENCE}
\newline
------------
\newline
The same target sentence inside its original block of text:
\newline
------------
\newline
\texttt{SENTENCE\_IN\_CONTEXT}
\newline
------------
\newline

**Question**:
Is the target sentence compliant? (YES/NO)

\# Answer: Lets think step-by-step. In order to provide the correct answer, you need to check if the target sentence matches the requirements. Provide the reasoning and the final answer (YES or NO) in \textless answer\textgreater  tags.

Reasoning:

\textless answer\textgreater YES/NO\textless/answer\textgreater

}
\end{tcolorbox}
\caption{Chain-of-Thoughts few-shot prompt template used for model experiments under the \emph{with-context} setup.}
\label{fig:appendix:prompt_template_with_context}
\end{figure*}

\begin{figure*}[p]
\begin{tcolorbox}[colback=gray!10!white, fontupper=\small, colframe=gray!80!black, title={CA-Judge Compliance Alignment Prompt}]
{\fontsize{7pt}{7pt}\selectfont
You are a senior compliance auditor of modern slavery statements with over 20 years of experience assessing corporate disclosures against modern slavery regulations for the Australian Modern Slavery Act. You possess expert knowledge of Australian Modern Slavery Act.\newline\newline
Your task is to evaluate the AI model's reasoning and its final answer based on a set of modern slavery assessment criteria. You will judge how well the AI reasoning aligns with these key rules and give the model reasoning a score based on whether the reasoning covers all the key rules and it follows the final answer.\newline\newline
\#\#\# EVALUATION DIMENSIONS:\newline
1. Accuracy - Identifies all relevant compliance gaps based on the key rules. Applies legal concepts exactly as defined — no misinterpretation or omission. No partial credit — any incorrect rule application makes the reasoning inaccurate.\newline
2. Clarity - Reasoning is logically structured with clear, step-by-step justification. Avoids vague terms, ambiguity, or unsupported claims. Final answer must clearly follow from the reasoning.\newline
3. Fidelity to Key Rules - All relevant key rules must be explicitly mentioned and addressed. Paraphrasing is allowed only if legal meaning is preserved. Irrelevant or external standards are penalized.\newline
4. Consistency - No internal contradictions; reasoning and conclusion must align.\newline
5. Evidence Use - Cites or paraphrases relevant rule clauses accurately. **No new rules shall be introduced.**\newline
6. Cognitive Behaviors (Verification \& Reflection) - Demonstrates explicit self-checking, cross-referencing, or reflection steps. Using phrases like "Wait, let me check", "I need to verify" is a must. This includes any self-correction or re-evaluation of the reasoning process while ensuring compliance with the key rules. **No new rules shall be introduced.** \newline

\#\#\# Scoring Rubric\newline\newline
| Score     | Description |\newline
|-----------|-------------|\newline
| **0.9-1.0**  | **Exceptional:** Demonstrates cognitive reasoning with explicit rule-by-rule cognitive behaviors, no omissions, and flawless logic. Final answer is directly and convincingly supported. Only awarded for perfect answers. |\newline
| **0.7-0.89** | **Strong, Near-Perfect:** Very good reasoning with clear logic and complete key rule assessment. Final answer must follow reasoning. Includes at least one Cognitive Behavior.|\newline
| **0.5-0.69** | **Good, but without Cognitive Behaviors:** Good reasoning with logic and mostly complete key rule application. Minor, non-critical omissions allowed. Final answer must follow reasoning. **But** shows no explicit verification or reflection.|\newline
| **0.3-0.49** | **Adequate but Flawed:** Reasoning shows effort but includes notable issues—missing key rules, partial logic, or weak justification. Final answer may be loosely connected to the reasoning. |\newline
| **0.1-0.29** | **Poor Reasoning:** Significant misunderstandings or misapplication of rules. Logic is unclear, unsupported, or misleading. Minimal evidence of comprehension. |\newline
| **0.0-0.09** | **Completely Incorrect:** No valid reasoning. Rules are ignored or misinterpreted. Final answer lacks justification or is based on fabricated logic. Even a correct guess receives 0.0. |\newline\newline

\#\#\# SCORING GUIDE\newline
Assign the scores based on the scoring rubric for each of the evaluation dimension and generate an overall score. The interval of scores in front of the scoring rubric indicates the scoring interval for this level, and you can score freely in this interval according to your confidence score. Be strict.\newline
No reasoning shall be given a score higher than 0.7 if there's no cognitive behavior. If the reasoning shows cognitive behaviors, add +0.05 but keep the total within the rubric ceiling.\newline
Noise Penalty: Deduct 0.05 from the overall score for each instance (or continuous cluster) of meaningless noise characters, random symbols, or stray HTML fragments found in the Reasoning or Final Answer.\newline

\#\#\# EVALUATION GUIDELINES:\newline
- Minor stylistic differences should not lower the score unless they affect the legal or ethical interpretation of the key rules.\newline
- The reasoning should consistently apply all the key rules without contradictions.\newline
- Flag reasoning that omits, distorts, or misinterprets critical compliance factors.\newline
- Ensure that the final answer is clearly justified by the reasoning provided.\newline
- Score the quality, not the length. If the reasoning is short but of high quality, it can still receive a high score.\newline
- Avoid scoring based on the final answer alone; focus on the reasoning's quality and its alignment with the key rules.\newline

\#\#\# INPUT FORMAT:\newline
- **Key Rules**: The set of key rules to comply a modern slavery criteria.\newline
- **Reasoning**: The model's reasoning process.\newline
- **Final Answer**: The model's final decision of whether the target sentence follows the key rules to match a criterion.\newline

\#\#\# REQUIRED OUTPUT FORMAT:\newline
Your response must include the following sections:\newline

**Reasoning**: Analyze how well the model's reasoning aligns with the provided modern slavery criteria key rules, discussing accuracy, clarity, and fidelity to the rules.\newline

**Score**: A score based on the Scoring rubric to evaluate the reasoning based on key rules. End with exactly this format:\newline
`The correctness score: [[score]]`\newline
The correctness score must strictly follow this format, e.g., `The correctness score: [[0.1]]`\newline

Do not provide additional text outside the required sections. Below are the key rules, reasoning and final answer you need to evaluate:\newline

\#\#\# Key Rules: \{RULES\}\newline

\#\#\# Reasoning: \{REASONING\}\newline

\#\#\# Final Answer: \{FINAL\_ANSWER\}\newline

\#\#\# The correctness score:\newline
---\newline
Based on the reasoning and the final answer, provide one correctness score in the correct format.\newline

}
\end{tcolorbox}
\caption{CA-Judge Compliance Alignment Prompt}
\label{fig:appendix:judge_prompt}
\end{figure*}

\section{Case study}
\label{ap:case_study}
We include a two-way case study that highlights how successive training stages improve the model’s reasoning and compliance performance. Figure \ref{tab:case-study-comparison-1} contrasts the outputs of the base model before and after fine-tuning (PT → FT). The fine-tuned model corrects the wrong prediction and provides more structured justifications, indicating that fine-tuning with format and correctness rewards substantially enhances reasoning capability. Figure \ref{tab:case-study-comparison-2} presents an example after integrating the rewards from CA-Judge. Relative to the FT variant, the CALLM delivers even richer step-wise rationales that explicitly cite the relevant key rules, further strengthening alignment with expert expectations.

\begin{figure*}[t]
\centering
\begin{tcolorbox}[colback=gray!5!white, colframe=gray!75!black, title={Case Study: Evaluating Model Outputs on Modern Slavery Compliance(C3, Risk Description)}]
\small
\textbf{Target Sentence to Classify:}
\small
\begin{quote}
Given the strong legislation and guidance in place, the risks of our operations being directly exposed to modern slavery is low
\end{quote}

\vspace{1em}
\textbf{Full Context:}

\small
\begin{quote}
who supply specialised consultancy and technical advice to Busways Group, are based in North America Risks of modern slavery practices in the operations and supply chains, and any entities the Business owns or controls 1.3.1 Risk in our operations From a geographical perspective, Australia is based in a low-risk region \textbf{Given the strong legislation and guidance in place, the risks of our operations being directly exposed to modern slavery is low} However, Busways Group is aware that there may be indirect exposure to risks further down our supply chain Busways Group has strong visibility over its own employees and complies with Australian workplace laws, including the Fair Work Act 2009 (Cth), National Employment Standards (NES), and work health and safety legislation
\end{quote}

\vspace{1em}
\textbf{Key rules:}

\small
\begin{quote}
 \textbf{Relevant sentences:} Relevant sentences may include descriptions such as: Describing specific modern slavery risk areas such as geographic regions (e.g., Indonesia), industries (e.g., electronics), commodities (e.g., palm oil). Sentences describing any past or current instances of modern slavery within the entity's operations or supply chains. Offering an explanation for why they believe that their risks of modern slavery are low.

 \textbf{Irrelevant sentences may include:} Hypothetical scenarios or vague claims like “Modern slavery might exist within the technology sector,” as well as broad definitions of risk that do not specifically relate to the organisation’s operations or supply chains or its own and controlled entities. Descriptions of other business risks (e.g., health, regulatory, environmental) without linking them to modern slavery. Merely stating that the company has zero or low risks is too vague; they must clarify why and on what analysis this is based. 
\end{quote}

\vspace{1em}
\textbf{Question:} Is the target sentence compliant? (YES/NO) \textbf{Ground Truth: YES}

\vspace{0.5em}
\begin{tabularx}{\textwidth}{@{}lX@{}}
\toprule
\textbf{Base Model (PT)} &
\small
The target sentence talks about low-risk levels for modern slavery practices due to strong legislation and guidance, as well as the cautious awareness towards indirect exposure through the supply chain. It does not provide specific risk areas, years of current or past cases within the entity's operations, or a clear explanation for why the risks are minimized. This meets the key rule for irrelevant sentences, which states that vague statements or general warnings about low risks without linking them to the entity's specific situation are not considered relevant.

\textbf{Answer: NO} \\
\midrule
\textbf{Base Model (FT)} &
\small
The target sentence provided states: "Given the strong legislation and guidance in place, the risks of our operations being directly exposed to modern slavery is low". This sentence directly ties to the Act’s requirement to report on modern slavery risks. \newline
Based on this sentence, the entity explicitly identifies a risk as low and explicitly attributes it and its risk to strong legislation and guidance. \newline
This explicitly links to a specific risk area (a low risk operation) relevant to the entity's operations, which is a critical feature to be identified as relevant under the Act.\newline
- Specifically, it describes a known past or present risk area ("Modern slavery is low") and a reason for this risk identification ("strong legislation and guidance").\newline
- This sentence is directly in the entity’s narrative and it is not just a hypothetical or general statement about future risks, but a direct declaration of a known low risk area.\newline
- It does not merely describe that the entity has certain "negative" or "positive" instances of past or current modern slavery, but explains a specific risk area of being low through strong legislation and guidance.\newline
- In addition, it specifies a risk area (direct exposure) and directly attributes it to a legal or regulatory provision, which aligns with the requirement for stating high-risk areas through specific entities or regulations which is relevant.\newline

\textbf{Answer: YES} \\
\end{tabularx}
\end{tcolorbox}
\caption{Case study comparing model responses on a compliance classification task, illustrating the effect of fine-tuning. The fine-tuned model (FT) rectifies errors made by the pre-trained model (PT) and produces better justifications.}
\label{tab:case-study-comparison-1}
\end{figure*}

\begin{figure*}[t]
\centering
\begin{tcolorbox}[colback=gray!5!white, colframe=gray!75!black, title={Case Study: Evaluating Model Outputs on Modern Slavery Compliance(C3, Risk Description)}]
\small
\textbf{Target Sentence to Classify:}
\small
\begin{quote}
As discussed in this Statement, we believe the nature of our business and supply chains mitigates modern slavery risks
\end{quote}
\vspace{0.5em}
\textbf{Full Context:}
\small
\begin{quote}
so for sale on our behalf, Red Hats supply chain is limited and consists of goods and services procured to operate our business and enable our employees and ecosystem of partners to deliver our services Red Hat procures goods and services from a diverse range of suppliers across several categories \textbf{As discussed in this Statement, we believe the nature of our business and supply chains mitigates modern slavery risks} Page 1 DocuSign Envelope ID: 32A5F4A3-1F99-4BA9-9A57-1AC6CB355564 Risks of Modern Slavery Practices in Our Operations and Supply Chain Operations We consider Red Hats risk of modern slavery in its operations to be relatively low, given our hiring practices and the contractual, policy, and statutory protections provided to all employees
\end{quote}

\textbf{Key rules:}
\small
\begin{quote}
 \textbf{Relevant sentences may include:} Describing specific modern slavery risk areas such as geographic regions (e.g., Indonesia), industries (e.g., electronics), commodities (e.g., palm oil). Sentences describing any past or current instances of modern slavery within the entity's operations or supply chains. Offering an explanation for why they believe that their risks of modern slavery are low.\newline
 \textbf{Irrelevant sentences may include:} Hypothetical scenarios or vague claims like “Modern slavery might exist within the technology sector,” as well as broad definitions of risk that do not specifically relate to the organisation’s operations or supply chains or its own and controlled entities. Descriptions of other business risks (e.g., health, regulatory, environmental) without linking them to modern slavery. Merely stating that the company has zero or low risks is too vague; they must clarify why and on what analysis this is based. 
\end{quote}
\textbf{Question:} Is the target sentence compliant? (YES/NO) \textbf{Ground Truth: YES}

\vspace{0.5em}
\begin{tabularx}{\textwidth}{@{}lX@{}}
\toprule
\textbf{Base Model (FT)} &
\small
Firstly, let's understand the target sentence: "As discussed in this Statement, we believe the nature of our business and supply chains mitigates modern slavery risks."
This sentence is stating "we believe" that our business and supply chains mitigate modern slavery risks, which aligns with the key rule that relevant sentences must explicitly describe modern slavery risks or mitigate them. 
However, the sentence does not provide any specific, detailed information like identifying geographic regions, industries, or supply chain elements that could lead to modern slavery. Instead, it just offers a generalized belief based on the nature of the company's operations.
Given that the primary requirement for a relevant sentence is to accurately describe modern slavery risks or mitigate them (without just stating a very vague belief), this sentence should be considered as irrelevant according to the provided key rules. It's a declaration of the belief but does not detail or substantiate the perceived mitigation of risks.

\textbf{Answer: NO} \\
\midrule
\textbf{CALLM} &
\small
The target sentence provided says: "As discussed in this Statement, we believe the nature of our business and supply chains mitigates modern slavery risks." This statement directly addresses the entity's risk areas of modern slavery. It specifies a general "risk mitigation" based on a factor (business and supply chains) which directly identifies an actionable risk area (low risk).\newline
1. The sentence explicitly links to the entity's business and supply chains as mitigating risks. \newline
2. This is a direct description of a specific area that might mitigate modern slavery risks (a known area that helps avoid risk areas, including geographic regions or industries known to be high risk) and it does not generalize or speculate about broader areas.\newline
3. It's also a fact-based assertion rather than a vague or hypothetical or stating general or broad risks that would be irrelevant to the entity's operations or supply chains.\newline
4. This statement, therefore, directly describes a specific modern slavery risk area (business and supply chains) to mitigate potential risks. \newline
5. The sentence also directly provides a basis for the entity's risk assessment and does not merely state or assume other business risks without linking to modern slavery specifically.\newline
6. Hence, the sentence is specific and directly applicable as described under Rule 1 of identifying that modern slavery risks are mitigated by certain business and supply chains.\newline
8. The sentence describes a past event (a concrete, actionable risk mitigation area) and thus also falls under Rule 1's criteria to provide explanation or mitigation of risk areas as a past event.

\textbf{Answer: YES} \\
\end{tabularx}
\end{tcolorbox}
\caption{Case study showing the additional benefit of CA-Judge rewards. Combining with CA-Judge yields more detailed, statute-referencing rationales and higher compliance accuracy. The CALLM demonstrates a more nuanced interpretation aligned with rule-based expectations.}
\label{tab:case-study-comparison-2}
\end{figure*}

\begin{figure*}[ht]
\begin{tcolorbox}[
    colback=gray!10!white,
    colframe=gray!80!black,
    boxrule=0.3pt,
    left=2pt,
    right=2pt,
    top=1pt,
    bottom=1pt,
    enhanced jigsaw,
    sharp corners,
     title={Case Study: CALLM Failure Case for the Approval criterion},
    fontupper=\fontsize{8pt}{12pt}\selectfont,
    boxsep=2pt
]
\textbf{Target Sentence to Classify:} \textit{The Board of}

\textbf{CALLM Reasoning:}
First, let's identify the target sentence and its context. The target sentence here alone: "The Board of Committee is formulating an education plan for external suppliers, through which we hope to raise awareness and create more collaboration across our supply chain Modern Slavery Statement FY22 Page 8 of 8 This Statement was approved by the boards of each of the three reporting entities covered by this Statement The Board of Bhaajayu Pty Ltd Busways Australia Pty Ltd Busways Blacktown Pty Ltd Busways Camden Pty Ltd Busways Campbelltown Pty Ltd Busways Gosford Pty Ltd Busways North West Pty Ltd Busways Pacific Pty Ltd Busways South Australia Pty Ltd Busways Wyong Pty Ltd Approved this statement on: 29 November 220" - there is no indication of joint approval or joint reporting statement. The sentence simply states that the board of each entity involved is formulating education plans and approving the statement, making it one body's declaration.

1. \textit{Verification for Direct Approval:} There is no mention of committee or other bodies involved in the approval of this statement.

2. \textit{Verification for Explicit Approval:} The target sentence is given without reference to committees. It clearly states "approved by the boards of each entity involved." This indicates that every entity involved's board had direct involvement in the statement approval.

3. \textit{Verification for Clarity:} The statement is clear that the approval authority is the principal governing body of each reporting entity involved. The sentences specifically highlight which entity's Board is involved: Bhaajayu Pty Ltd, Busways Australia Pty Ltd, Busways Blacktown Pty Ltd, Busways Camden Pty Ltd, Busways Campbelltown Pty Ltd, Busways Gosford Pty Ltd, Busways North West Pty Ltd, Busways Pacific Pty Ltd, Busways South Australia Pty Ltd, Busways Wyong Pty Ltd.

\textbf{CALLM Final Answer:} \textit{no} (Ground truth: yes)

\textbf{CA-Judge response:}

1. \textit{Accuracy:}
   - The reasoning correctly identifies that the target sentence does not mention any committee or delegated approval authority, adhering to the rule that such delegations are explicitly prohibited.
   - However, the statement does mention the approval coming from the boards of each individual entity, which could be considered a direct approval by the principal governing body. The reasoning fails to recognize this as it states "indirect approval," which might imply a lack of direct involvement from the principal governing body.
   - The reasoning also mentions that there is no confusion regarding which governing board approved the statement, but it incorrectly concludes that this constitutes "indirect approval."

2. \textit{Clarity:}
   - The reasoning is somewhat clear in identifying the absence of committee or delegated approval. However, it could be more precise in distinguishing between direct and indirect approval.
   - The final statement, "All these verification points demonstrate that the statement explicitly states indirect approval by a governing body," is ambiguous and potentially misleading since it does not clearly differentiate between direct and indirect approval.

3. \textit{Fidelity to Key Rules:}
   - The reasoning does not fully address the key rule that approval must come directly from the principal governing body. Instead, it suggests that the approval is "indirect," which might be a misinterpretation.
   - The statement "approved by the boards of each entity involved" should be interpreted as direct approval, thus failing to meet the requirement that the approval must come from the principal governing body (Board of Directors).

4. \textit{Consistency:}
   - There is a contradiction in the reasoning where it initially suggests that the approval is indirect, but then clarifies that the statement is clear about the governing board's involvement. This inconsistency might lead to confusion.

5. \textit{Evidence Use:}
   - The reasoning correctly cites the specific boards involved in the approval, which is accurate evidence.
   - However, the interpretation of "indirect approval" is not aligned with the key rule requiring direct approval from the principal governing body.

6. \textit{Cognitive Behaviors:}
   - The reasoning does not explicitly verify or reflect on the potential misinterpretation of "indirect approval" versus "direct approval."

\textit{Conclusion:} Given the misinterpretation of "indirect approval" as not meeting the requirement of direct approval from the principal governing body, the reasoning is flawed. Therefore, the final answer "NO" is correct, but the reasoning provided does not fully support this conclusion accurately.
\textit{Score: [[0.6]]}

\end{tcolorbox}
\caption{
Illustrative Use Case of CALLM with CA-Judge Scoring. In the example, CALLM incorrectly predicts non-compliance for the Approval criterion. This error can be traced to persistent parsing issues in the dataset stemming from PDF scraping artifacts. We can also see incorrect parsing of a table causing repeated entities mentioned in the sentence. CA-Judge accurately identifies flaws in the model’s reasoning and assigns a lower score, consistent with the rule violations defined in Figure 1 in paper. 
}
\label{fig:model_output_failure_1}
\end{figure*}

\bibliography{aaai2026}

\end{document}